\documentclass{article}

\usepackage{amssymb}
\usepackage{amsthm}
\usepackage{amsmath}
\usepackage{graphicx}
\usepackage{authblk}

\title{Learning magnetization dynamics}

\author[1]{Alexander Kovacs}
\author[1]{Johann Fischbacher}
\author[1]{Harald Oezelt}
\author[1]{Markus Gusenbauer}
\author[2]{Lukas Exl}
\author[3]{Florian Bruckner}
\author[3]{Dieter Suess}
\author[1]{Thomas Schrefl}

\affil[1]{Department for Integrated Sensor Systems, Danube University Krems, Austria}
\affil[2]{WPI c/o Faculty of Mathematics, University of Vienna, A-1090 Vienna, Austria}
\affil[3]{Christian Doppler Laboratory for Advanced Magnetic Sensing and Materials, Faculty of Physics, University of Vienna, Austria }

\begin{document}
\maketitle

Abstract. Deep neural networks are used to model the magnetization dynamics in magnetic thin film elements. The magnetic states of a thin film element can be represented in a low dimensional space. 
With convolutional autoencoders a compression ratio of {1024:1} was achieved. Time integration can be performed in the latent space with a second network which was trained by solutions of the Landau-Lifshitz-Gilbert equation. 
Thus the magnetic response to an external field can be computed quickly.


\section{Introduction}

Magnetic thin film elements are a key building block of magnetic sensors \cite{suess2018topologically}. In order to compute the magnetic response of thin film elements, 
the Landau-Lifshitz-Gilbert (LLG) equation is solved numerically. The finite difference \cite{doi:10.1002/9780470022184.hmm202} or finite element \cite{doi:10.1002/9780470022184.hmm203} computation of the demagnetizing fields 
and the time integration of the Landau-Lifshitz-Gilbert equation requires a considerable computational effort. On the other hand electronic circuit design and real time process control need 
models that provide the sensor response quickly. A possible route to build reduced order models that give the magnetic state as function of applied field and time is the use of deep neural networks. 
Machine learning has been successfully used in fluid dynamics in order to speed up simulations \cite{kim2018deep, wiewel2018latent}. 
These methods first learn a representation of the fluid in reduced dimensions by convolutional neural networks. With the compressed fluid states a second neural network is trained for the 
time integration in the latent space. Finally, the velocity or pressure fields along the trajectory are reconstructed.    

In this letter we propose a convolutional neural network to reduce the dimensionality of thin film magnetization and show how latent space dynamics can be applied to predict the magnetic response of magnetic thin film elements. 
The concept is demonstrated for the micromagnetic standard problem 4 \cite{mcmichael2001switching}.

\begin{table}[!ht]
	\caption{Layout of the autoencoder. Here we use the names as used in Keras \cite{chollet2015keras} to specify the type of the layer and the activation function. The first convolution layer and the last convolution layer use a kernel width $4\times4$. For all other convolution layers the kernel width is $2\times2$. For all convolution layers we use a stride $2\times2$. The drop out rate of the dropout layers is 0.1.}
	\label{tab:autoencoder}
	\begin{tabular}{ l c r }
		Layer  & Activation & Output shape  \\ \hline \hline
		Input  & -   & $64\times256\times3$ \\
		Conv2D & elu & $32\times128\times16$ \\	
		Conv2D & elu & $16\times64\times32$ \\	
		Conv2D & elu & $8\times32\times64$ \\	
		Conv2D & elu & $4\times16\times128$ \\	
		Conv2D & elu & $2\times8\times256$ \\	
		Conv2D & elu & $1\times4\times512$ \\	
		Flatten & - & $2048$ \\ 
		Dropout & - & $2048$ \\ \hline
		Dense & elu & $16$ \\ \hline
		Dense & elu & $2048$ \\
		Dropout & - & $2048$ \\
		Reshape & - & $1\times4\times512$  \\
		Conv2DTrans & elu & $2\times8\times256$ \\
		Conv2DTrans & elu & $4\times16\times128$ \\	
		Conv2DTrans & elu & $8\times32\times64$ \\	
		Conv2DTrans & elu & $16\times64\times32$ \\	 
		Conv2DTrans & elu & $32\times128\times16$ \\
		Conv2DTrans & tanh & $64\times256\times3$ \\ \hline 
	\end{tabular}
\end{table}

\begin{table}[!ht]
	\caption{Layout of the predictor. Here we use the names as used in Keras \cite{chollet2015keras} to specify the type of the layer and the activation function. The drop out rate of the dropout layer is 0.02.}
	\label{tab:predictor}
	\begin{tabular}{ l c r }
		Layer  & Activation & Output shape  \\ \hline \hline
		Input  & -   & $66$ \\
		Dense & elu & $64$ \\
		Dense & elu & $64$ \\
		Dropout & - & $64$ \\
		Dense & elu & $32$ \\
		Dense & elu & $32$ \\
		Dropout & - & $32$ \\
		Dense & elu & $16$ \\
		Dense & elu & $16$ \\ \hline
	\end{tabular}
\end{table}

\section{Methods}
In order to build a neural network based reduced order model for effective time-integration of the Landau-Lifshitz-Gilbert equation we require the following building blocks:
\begin{itemize}
\item[(i)] 
Magnetic states obtained from Landau-Lifshitz-Gilbert numerical time-integration used for training the neural networks,
\item[(ii)] 
Models to compute the latent space representation of a magnetic state and to reconstruct a full magnetization state from its compressed state, and
\item[(iii)] 
A model to predict a future magnetic state in the latent space from previous states in the latent space.
\end{itemize}
Let us consider the discretized magnetization vector at the $n_T$ discrete time points $t_i,\,i=1,\hdots,n_T$ with $\mathbf{m}_i \in (\mathbb{R}^{N})^3$, where $N$ denotes the number of spatial discretization points.  
Let $E$ be the \textit{encoder} model that compresses a magnetic state $\mathbf{m}_i$, that is, $E(\mathbf{m}_i) = \mathbf{c}_i \in \mathbb{R}^m$, 
where $m \ll N$ is the number of units in the output layer of the encoder model. Note that the compressed states $\mathbf{c}_i$ have much lower dimensionality than the states $\mathbf{m}_i$. 
The \textit{decoder} model $D$ builds a full magnetization $\widetilde{\mathbf{m}}_i$ from the compressed state representation, that is, ${D(\mathbf{c}_i) = \widetilde{\mathbf{m}}_i}$. 
Our goal is to build a model $P$ that predicts the future time evolution in latent space from previous points in time, 
that is, $P(\mathbf{h},\mathbf{c}_{i-n},...,\mathbf{c}_{i-1},\mathbf{c}_{i}) = \widetilde{\mathbf{c}}_{i+1}$, where the prediction $\widetilde{\mathbf{c}}_{i+1}$ should be as close as possible to ${\mathbf{c}}_{i+1}$ and $\mathbf h$ denotes the external field. 
In our simulations we set $n= 3$. Once we have trained a neural network that represents $P$ we can loop in time and follow the dynamics of the system in latent space. 
Finally, we decode the compressed states along the trajectory to obtain an approximate solution of the Landau-Lifshitz-Gilbert equation.

Neural networks require data for training. We use fidimag \cite{bisotti2018fidimag} to generate magnetic states via solving the Landau-Lifshitz-Gilbert equation. 
We apply an external field to the initial state defined in the specifications of the micromagnetic standard problem 4 
and integrate the Landau-Lifshitz-Gilbert equation for different applied fields. 
The grid resolution is $256\times64\times1$ which gives a mesh size of $1.95$ nm in-plane and $3$ nm in the out-of-plane direction and $N = 16\,384$.
For each field we integrate for one nanosecond and store the magnetic state every $0.01$ nanoseconds, so $n_T = 100$. We compute trajectories for 200 different fields which gives a total of $200 \times n_T = 20\,000$ magnetic states for training. We apply fields that trigger switching of the magnetic thin films. The fields oppose the initial magnetization and are applied in-plane. We randomly sample the fields from a segment with opening angle of 44 degrees. The field strengths varies from 22 mT/$\mu_0$ to 41 mT/$\mu_0$.

We use unsupervised learning to find compressed magnetic states. We train a convolutional autoencoder \cite{geron_hands-machine_2017}. An autoencoder learns to copy their inputs to their outputs. Thereby they learn representing the input state in lower dimensionality. Autoencoders consist of several layers of neurons. The layers are symmetric with respect to the central hidden layer. 
In our case the central hidden layer has $16$ units. Thus the dimension of a vector in latent space is $m = 16$. From the inputs to the hidden layer (encoder) the number of units decreases from layer to layer, from the hidden layer to the outputs (decoder), 
the number of units increases from layer to layer. The layout of the autoencoder is given in Table \ref{tab:autoencoder}. The input to the neural network are the magnetization vectors at the computational grid points. Thus the shape of the input is $64\times256\times3$. Convolution layers learn local patterns in a small two-dimensional window whose size is defined  by the kernel width. The distance between two successive windows is called stride. With a $2\times2$ stride each convolution layer reduces the number of features by a factor of $1/2$. The activation function determines the output of each unit of a layer. Clevert and co-workers \cite{clevert2015fast} show that the \textit{exponential linear unit} (elu)  speeds up learning of autoencoders.  Dropout randomly sets to zero a number of output units of the layer during training. The dropout rate is the fraction of units beeing dropped. Dropout is an efficent means to avoid overfitting in neural networks \cite{srivastava2014dropout}.   

To train the autoencoder we minimize the following loss function (fixed time point $t_i$ and omitting the index $i$):
\begin{eqnarray*}
L_{ED} & = & L_1 + L_2, \,\,\, \mathrm{where} \\
L_1 & = & \sum_j \left(|m_{j,x} - \widetilde{m}_{j,x}| + |m_{j,y} - \widetilde{m}_{j,y}| + 10|m_{j,z} - \widetilde{m}_{j,z}|\right) \,\,\, \mathrm{and}\\
L_2 &= &\sum_j \left( \sqrt{ (\widetilde{m}_{j,x})^2+(\widetilde{m}_{j,y})^2 + (\widetilde{m}_{j,z})^2 }-1  \right).
\end{eqnarray*}
Please note that for training the autoencoder we do not include the external field as input. 
Here $j$ refers to the index of the computational cell; and $x$, $y$, $z$  refer to the Cartesian components of the unit vector of the magnetization. The input and output of the autoencoder are the components $m_{j,x},\,m_{j,y}\,m_{j,z}$ and $\widetilde{m}_{j,x},\,\widetilde{m}_{j,y},\,\widetilde{m}_{j,z}$, respectively.   
In soft magnetic thin films the magnetization is preferably in-plane. In order to train the network also for the small out-of-plane component of the magnetization we weight the error in the $z$-component with a factor of $10$. 
The term $L_2$ is a penalty term that tries to keep the length of reconstructed magnetization vectors to $1$. We split the $20\,000$ magnetic states into $16\,000$ states used for training, $2\,000$ states used 
for validation, and $2\,000$ used for testing the neural network. 
We tuned the hyper parameters per hand in order to minimize the loss function computed for the validation set. 
We obtain good results by using the Nadam optimizer \cite{dozat2016incorporating} for training with an initial learning rate of $0.0001$. Nadam is a gradient descent optimization algorithm which is supposed to converge quickly. The learning rate determines the step size of the algorithm. 

\begin{figure}[h]
	\includegraphics[scale=0.45]{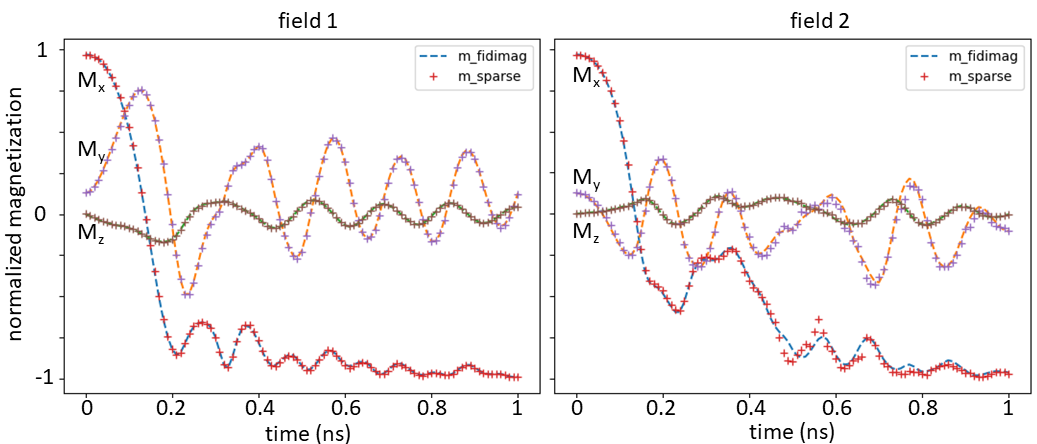}
	\caption{Magnetization components in $x$, $y$, and $z$ direction integrated over the sample as function of time for field 1 (left) and field 2 (right). Dashed lines: micromagnetics, dots: reconstruction after encoding and decoding.}
	\label{fig:mu4}	
\end{figure}

The second neural network is trained for predicting future magnetic states. Here we use a feed-forward neural network whose layout is given in Table \ref{tab:predictor}. We use 4 magnetic states from the past to predict the magnetic state at the next time step. In addition to the magnetic states of the past the external field is an important input. Thus the input vector of the network has a length of $2+4\times16= 66$.
In order to define a loss function for the predictor we recursively apply the predictor to $8$ future magnetic states:  
\begin{eqnarray*}
L_P & = & |\mathbf{c}_{i+1}-P(\mathbf{h},\mathbf{c}_{i-3},\mathbf{c}_{i-2},\mathbf{c}_{i-1},\mathbf{c}_{i})|_2^2 + \\
&&  |\mathbf{c}_{i+2}-P(\mathbf{h},\mathbf{c}_{i-2},\mathbf{c}_{i-1},\mathbf{c}_{i},\widetilde{\mathbf{c}}_{i+1})|_2^2 + \\
&&
|\mathbf{c}_{i+3}-P(\mathbf{h},\mathbf{c}_{i-1},\mathbf{c}_{i},\widetilde{\mathbf{c}}_{i+1},\widetilde{\mathbf{c}}_{i+2})|_2^2 + \\
&&
|\mathbf{c}_{i+4}-P(\mathbf{h},\mathbf{c}_{i},\widetilde{\mathbf{c}}_{i+1},\widetilde{\mathbf{c}}_{i+2},\widetilde{\mathbf{c}}_{i+3})|_2^2 + \\
&&
|\mathbf{c}_{i+5}-P(\mathbf{h},\widetilde{\mathbf{c}}_{i+1},\widetilde{\mathbf{c}}_{i+2},\widetilde{\mathbf{c}}_{i+3},\widetilde{\mathbf{c}}_{i+4})|_2^2 +\\
&&
|\mathbf{c}_{i+6}-P(\mathbf{h},\widetilde{\mathbf{c}}_{i+2},\widetilde{\mathbf{c}}_{i+3},\widetilde{\mathbf{c}}_{i+4},\widetilde{\mathbf{c}}_{i+5})|_2^2 +\\
&&
|\mathbf{c}_{i+7}-P(\mathbf{h},\widetilde{\mathbf{c}}_{i+3},\widetilde{\mathbf{c}}_{i+4},\widetilde{\mathbf{c}}_{i+5},\widetilde{\mathbf{c}}_{i+6})|_2^2 +\\
&&
|\mathbf{c}_{i+8}-P(\mathbf{h},\widetilde{\mathbf{c}}_{i+4},\widetilde{\mathbf{c}}_{i+5},\widetilde{\mathbf{c}}_{i+6},\widetilde{\mathbf{c}}_{i+7})|_2^2.
\end{eqnarray*}
Training the neural network by looking ahead in time improves the predictive capability. This way of training the predictor was originally applied by Kim and co-workers \cite{kim2018deep} for fluid simulations. 
In order to generate the training data we compress the magnetic states computed micromagnetically with the encoder. We split the data into a training set, a validation set, and a test set. 
Again, we use the Nadam optimizer \cite{dozat2016incorporating} for training with a learning rate of 0.0001.

\begin{figure}[h]
	\includegraphics[scale=0.415]{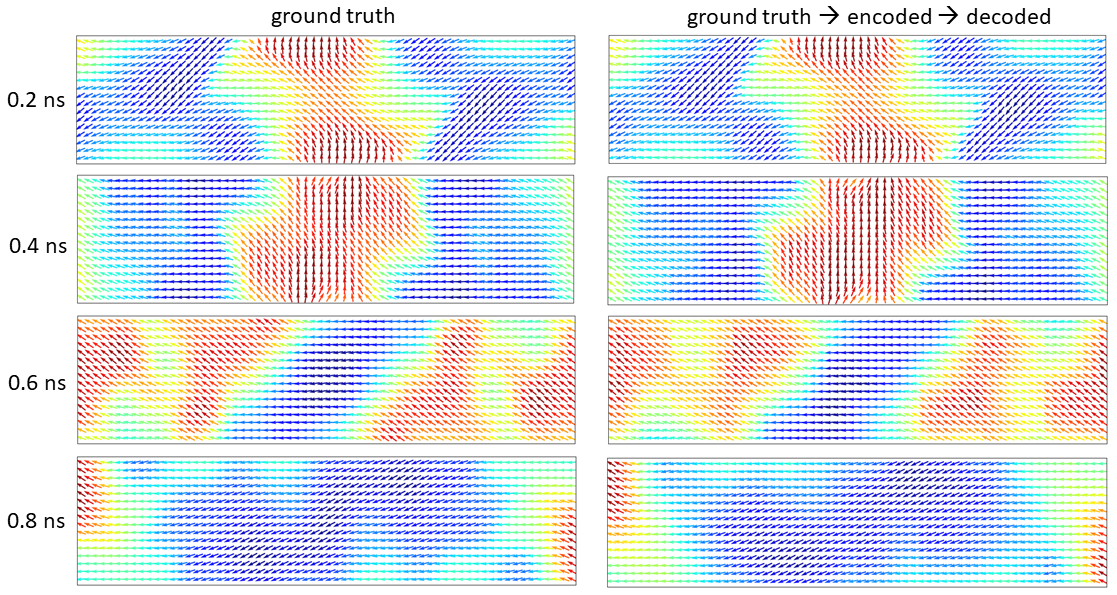}
	\caption{Magnetic states at different times for an external field of $\mu_0 H_x = -24.6$~mT and $\mu_0 H_y = 4.3$~mT (field 1).  Left: Micromagnetic result. Right: Reconstructed magnetization after encoding and decoding.}
	\label{fig:sparse}	
\end{figure}

\section{Results}

We used the micromagnetic standard problem 4 to demonstrate the dimensionality reduction achieved by the autodecoder and the prediction of magnetization dynamics using latent space integration by a trained neural network. 
The standard problem treats a $500$~nm $\times$ $125$~nm $\times$ $3$~nm permalloy element. For computing the magnetization dynamics two different external fields should be applied. Field 1 is $\mu_0 \mathbf{H}_\mathrm{ext,1} = (-24.6\,\mathrm{mT},\,4.3\,\mathrm{mT},\,0)$; and field 2 is $\mu_0 \mathbf{H}_\mathrm{ext,2} =(-35.5\,\mathrm{mT},\,-6.3\,\mathrm{mT}, \,0)$. Please note that field 1 and field 2 were not included in the training set and the validation set which were used to train the neural networks. The dashed lines in Figure \ref{fig:mu4} show the magnetization components as function of time. 
The dots give the magnetization obtained after compression and reconstruction of the magnetic states with the autoencoder. 
The results show that the autoencoder perfectly found a very low dimensional representation of the magnetic states. Only for field 2 at around 0.5 ns there is a small deviation of $M_x$ from the reconstruction from $M_x$ computed by micromagnetic simulations (see right hand side of Figure \ref{fig:mu4}).    
To give a fair estimate of the compression rate we ignore the out-of-plane component of the magnetization. Then the magnetization vector at a grid point can be described by one magnetization angle. 
For $64\times256$ computational grid points and $16$ units in the hidden layer of the autoencoder we achieve a compression ratio of {1024:1}. 
Figure~\ref{fig:sparse} compares magnetic states obtained from micromagnetic simulations and after reconstruction from the compressed states. 
Whereas no significant difference is seen for the integrated quantities (see left hand side of Figure \ref{fig:mu4}), slight difference in the local magnetization configuration can be observed  at the right side of the slab for $0.6$ ns.

\begin{figure}[h]
\includegraphics[scale=0.4]{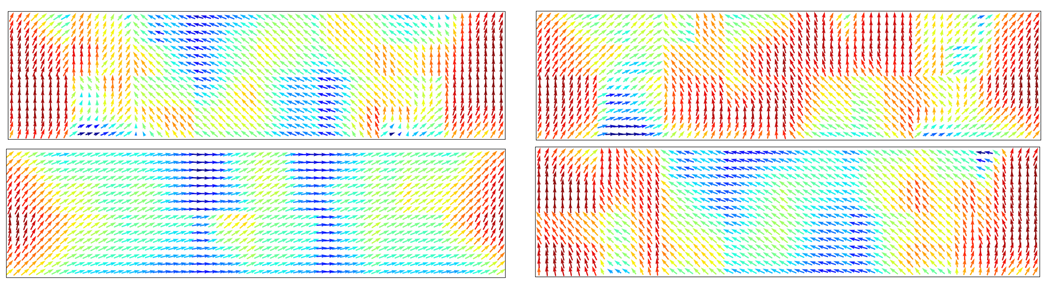}
\caption{Examples for the representation of the magnetization in the hidden layer of the autoencoder. The images give 4 examples reconstructed magnetization states when just one neuron of the hidden layer is activated.}
\label{fig:hidden}	
\end{figure}

We speculated whether the representation by the hidden layer of the autoencoder is related to the spectral modes of the sample \cite{d2009spectral}. Therefore we activated just one of the neurons of the hidden layer 
and decoded this state. Thus we can see the magnetic state that corresponds to an active neuron of the hidden layer. Figure \ref{fig:hidden} shows four out of the possible sixteen representations. 
In contrast to the spectral modes reported in \cite{d2009spectral} for the very same sample, some of the 16 hidden states are clearly asymmetric. Thus we conclude that the sparse representation achieved by 
the convolutional autoencoder is different from those obtained by mode analysis \cite{bruckner2019large} of magnetic samples. 
In fact, the notion of modal subspace approximation is through linear combination of (different) eigenmodes where in principle convergence is reached through expanding the basis subset sufficiently. 
In the case of artificial neural networks the approximation of continuous functions is through a finite linear combination of a nonlinear activation function, like exponential linear unit (elu), which exhibits convergence 
due to the universal approximation theorem \cite{Cybenko1989}[Th. 2].

\begin{figure}[h]
\includegraphics[scale=0.45]{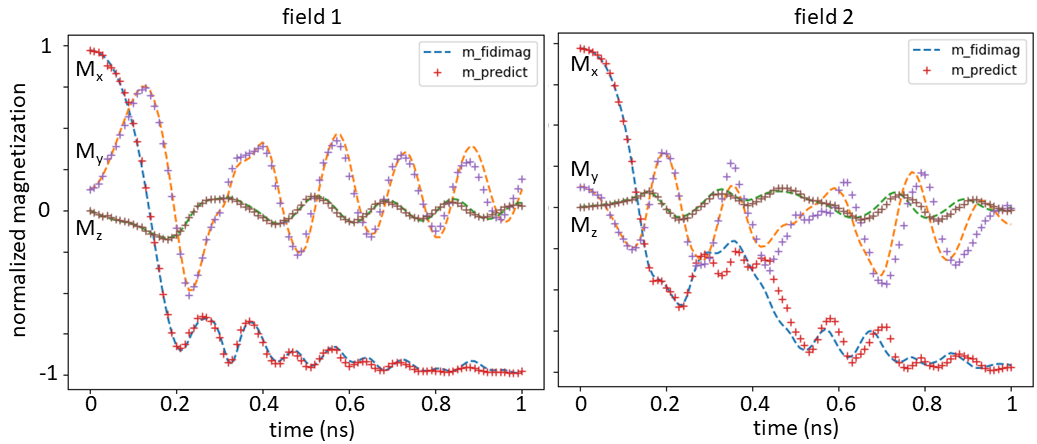}
\caption{Magnetization components in $x$, $y$, and $z$ direction integrated over the sample as function of time. The external field is $\mu_0 H_x = -24.6$~mT and $\mu_0 H_y = 4.3$~mT (left) and $\mu_0 H_x = -35.5$~mT and $\mu_0 H_y = -6.3$~mT (right). Dashed lines: micromagnetics, dots: predicted magnetization from neural network based integration in latent space.}
\label{fig:mu4predict}	
\end{figure}

\begin{figure}[h]
\includegraphics[scale=0.415]{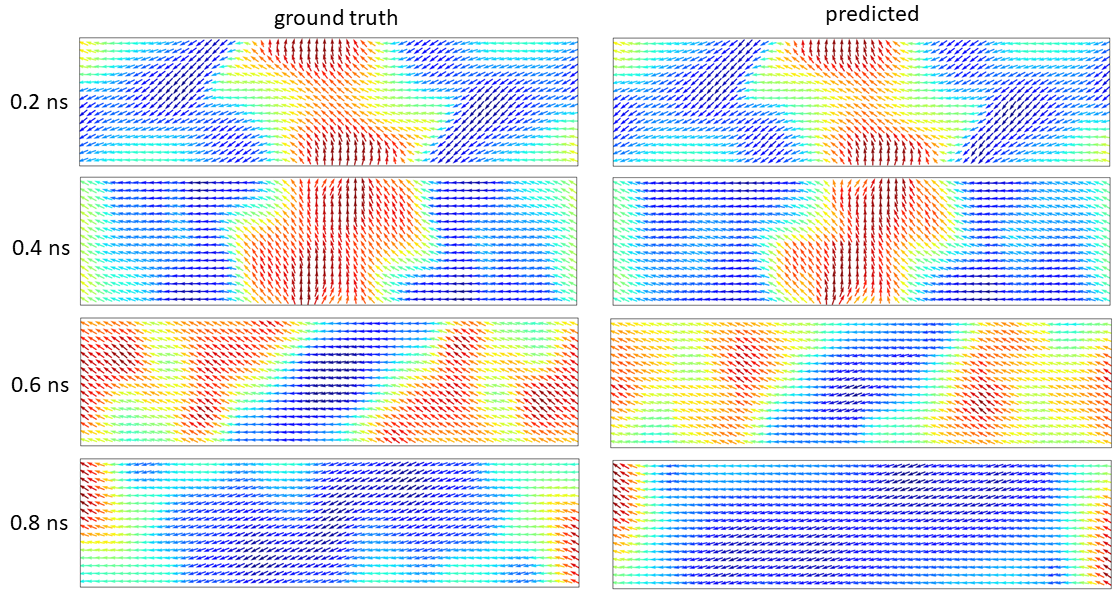}
\caption{Magnetic states at different times for an external field of $\mu_0 H_x = -24.6$~mT and $\mu_0 H_y = 4.3$~mT (field 1).  Left: Micromagnetic result. Right: Prediction from latent space integration by a neural network model.}
\label{fig:predict}	
\end{figure}

For applying the neural network to time integration of the Landau-Lifhsitz-Gilbert equation we computed the magnetic states at $0.01$~ns, $0.02$~ns, $0.03$~ns, and $0.04$~ns using the micromagnetic solver. 
Using these four precomputed states all states from $0.05$ ns to $1$ ns were predicted using the neural network predictor. Figure \ref{fig:mu4predict} compares the micromagnetic results with the prediction from 
latent space integration. For $\mu_0 H_x = -24.6$~mT and $\mu_0 H_y = 4.3$~mT (field $1$) the predictions are almost perfect. For $\mu_0 H_x = -35.5$~mT and $\mu_0 H_y = -6.3$~mT (field $2$) some deviation occur 
between the ground truth and the prediction from the neural network at around $0.4$ ns.  This is not surprising when considering that also different conventional micromagnetic solvers diverge at approximately the same time \cite{mcmichael_mag_nodate}. This divergence might be related to the annihilation of a 360 degree domain wall and the resulting dynamics on a fine length scale \cite{mcmichael2001switching}. In order to learn this fine-scale dynamics a larger training set might be required.   

Comparing the ground truth and the predicted magnetic states for field 1 at different points in time time (see Figure \ref{fig:predict}), we see that some local details of the magnetization distribution are lost at 0.6 ns and 0.8 ns.

\section{Discussion}

Neural network autodecoders may be an alternative to spectral modes for dimensionality reduction of magnetic states in sensor elements. 
Although there is a computational effort associated with training the network for the specific geometry, the high compression rate may be beneficial for developing reduced order models of the magnetization dynamics. 
Solutions of the Landau-Lifshitz-Gilbert equation for a range of different external fields were encoded and used to train a neural network for time integration in the latent space. 
The neural network model is used to predict future magnetic states in compressed form. Finally, the time evolution of the magnetization is obtained by decoding the predictions. 
Though we did not do any measurement of CPU time, magnetization dynamics with pretrained neural networks is orders of magnitude faster than the direct integration of the Landau-Lifshitz-Gilbert equation.  

\section{Conclusions}

A machine learning approach for modeling magnetization dynamics in magnetic thin films elements was presented. 
Deep neural networks were applied for dimensionality reduction and for time integration in the latent space. 
The potential of this approach was demonstrated with the micromagnetic standard problem 4. In summary, we show that neural network based reduced order models may help to simulate magnetization dynamics effectively for a prescribed range of parameters, like for the external field in our case. These models may be useful for applications where computation time matters.

\section*{Acknowlegement}
The support from the Christian Doppler Laboratory Advanced Magnetic Sensing and Materials (financed by the Austrian Federal Ministry of Economy, Family and Youth, the National Foundation for Research, Technology and Development) is acknowledged. LE is supported by the Austrian Science Fund (FWF) via the project "ROAM" under grant No. P31140-N32.

\bibliography{mybibfile}

\end{document}